\documentclass{aa}
\usepackage{graphicx}
\newcommand{\Msun}{\mbox{\,$M_{\odot}$}}        
\newcommand{\Rsun}{\mbox{\,$R_{\odot}$}}        
\newcommand{\kms}{\mbox{\,km\,s$^{-1}$}}        
\begin{document}

   \title{HS~0705+6700: a new eclipsing sdB binary\thanks{
          Based on observations collected at the German-Spanish 
          Astronomical Center (DSAZ), Calar Alto, operated by the
          Max-Planck-Institut f\"ur Astronomie Heidelberg jointly
          with the Spanish National Commission for Astronomy
         }$^,$\thanks{
          Based on observations obtained at the Nordic Optical 
          Telescope, operated on the island of La Palma jointly by
          Denmark, Finland, Iceland, Norway, and Sweden, in the Spanish
          Observatorio del Roque de los Muchachos of the Instituto de
          Astrofisica de Canarias}
         }

   \author{H. Drechsel\inst{1}
   \and    U. Heber\inst{1}
   \and    R. Napiwotzki\inst{1}
   \and    R. {\O}stensen\inst{2,3}
   \and    J.-E. Solheim\inst{3}
   \and    F. Johannessen\inst{3}
   \and    S. L. Schuh\inst{4}
   \and    J. Deetjen\inst{4}
   \and    S. Zola\inst{5}
          }

   \offprints{H. Drechsel,\\
              drechsel@sternwarte.uni-erlangen.de}

   \institute{Dr. Remeis-Sternwarte Bamberg,
              Astronomisches Institut der Universit\"at
              Erlangen-N\"urnberg, Sternwartstra\ss e 7,
              D-96049 Bamberg, Germany\\
              e-mail: {\tt drechsel@sternwarte.uni-erlangen.de}
   \and       Isaac Newton Group of Telescopes, E-37800 Santa Cruz de La Palma,
              Canary Islands, Spain\\
              e-mail: {\tt roy@ing.iac.es}
   \and       University of Troms\o, Department of Physics,
              N-9037 Troms\o, Norway\\
              e-mail: {\tt janerik@phys.uit.no}
   \and       Institut f\"ur Astronomie und Astrophysik,
              Universit\"at T\"ubingen,
              Waldh\"auser Stra\ss e 64, 72076 T\"ubingen, Germany\\
              e-mail: {\tt schuh@astro.uni-tuebingen.de}
   \and       Astronomical Observatory of the Jagiellonian University,
              ul. Orla 171, 30-244 Cracow, Poland\\
              e-mail: {\tt sfzola@cyf-kr.edu.pl}
             }

   \date{Received date / accepted date}

   \abstract{
We report the discovery of an eclipsing binary -- \object{HS~0705+6700} --
being an sdB star with a faint companion.
From its light curve the orbital period of 8263.87\,s, the mass ratio of the
system $q = 0.28$, the inclination of 84$\fdg4$ and other system parameters
are derived. The companion does not contribute to the optical light of the
system except through a strong reflection effect. The semi-amplitude of the
radial velocity curve $K_1 = 85.8$\kms\ and a mass function of
$f(m) = 0.00626$\Msun\ are determined. A spectroscopic analysis of the blue
spectra results in T$_{\rm eff} = 28\,800$\,K, $\log~g = 5.40$, and 
$\log (n_{\rm He}/n_{\rm H}) = -2.68$. These characteristics are typical for 
sdB stars, as is its mass of 0.48\Msun. According to its mass (0.13\Msun) and
radius (0.19\Rsun), the companion is an M dwarf. The primary is in a core
helium burning phase of evolution, and the system must have gone through a
common envelope stage when the primary was near the tip of the red giant
branch.
}

   \markboth{H. Drechsel et al.: HS~0705+6700: a new eclipsing sdB binary}
            {H. Drechsel et al.: HS~0705+6700: a new eclipsing sdB binary}

   \maketitle

\section{Introduction}

Subluminous B stars (sdB) dominate the populations of faint blue stars of
our own Galaxy and are found in both the old disk and in halo populations,
e.g. as stars forming the blue tails to the horizontal branches of globular
clusters (Ferraro et al.~\cite{ferraro97}).

In the context of galaxy evolution sdB stars are important because they are
sufficiently common to be the dominant source for the ``UV upturn
phenomenon'' observed in elliptical galaxies and galaxy bulges 
(Brown et al.~\cite{brown97}, \cite{brown00a}; Greggio \& Renzini 
\cite{greggio90}, \cite{greggio99}). They might also be used as age indicators
for elliptical galaxies (Brown et al.~\cite{brown00b}).

 
However, important questions remain over the evolutionary paths and the
appropriate timescales. There is general consensus that the sdB stars can be
identified with models for Extreme Horizontal Branch (EHB) stars
(Heber \cite{heber86}, Saffer et al.~\cite{saffer94}). Like all HB stars they
are core helium burning objects. However, their internal structure differs
from typical HB stars, because their hydrogen envelope is very thin
($<$1\% by mass) and therefore inert. As a consequence EHB stars evolve
directly to the white dwarf cooling sequence thus avoiding a second red giant
phase (Dorman et al.~\cite{dorman93}).


The discovery of multi-mode, short-period ($P = 2-10$ min.) pulsators 
among sdB stars (see O'Donoghue et al.~\cite{odono99} for a review) has
opened a new attracting possibility of probing their interiors using
seismological tools. Recently, Brassard et al.~(\cite{brassard01}) were able
to derive the mass and hydrogen envelope mass of the pulsating sdB star
\object{PG~0014+067} by asteroseismology. Their results (total mass
$M = 0.49 \pm 0.019$\Msun\ and envelope mass
$\log (M_{\rm env}/\Msun) = -4.3 \pm 0.22$) are in excellent agreement
with the predictions by evolution theory.

Considerable evidence is accumulating that many sdB stars reside in close 
binaries (Maxted et al.~\cite{maxted01a}, Saffer et al.~\cite{saffer01}). 
Therefore mass transfer should play an important role in the evolution
of such binary systems. Detailed investigations of sdB binaries, in
particular eclipsing systems, are crucial to determine their masses.
However, only two such eclipsing binaries, \object{HW~Vir} (Menzies \& Marang
\cite{menzies86}) and \object{PG~1336-018} (Kilkenny et al.~\cite{kilkenny98})
are known up to now, which consist of an sdB star and an optically invisible
M dwarf companion. Here we report the discovery of the third such system --
\object{HS~0705+6700}. 
 
\object{HS~0705+6700} was selected from a list of candidates drawn from the
Hamburg Schmidt survey (Hagen et al.~\cite{hagen95}). Follow-up spectroscopy
(Heber et al.~\cite{heber99}, Edelmann et al.~\cite{edelmann01}) revealed 
that its effective temperature lies in the domain predicted for the 
pulsational instability. Therefore, \object{HS~0705+6700} was included in a 
photometric monitoring programme at the Nordic Optical Telescope 
(see {\O}stensen et al.~\cite{ost01a}, \cite{ost01b}) in order to search for 
pulsations. Indeed, the star turned out to be variable, but not on the
time and flux scale expected for pulsations.
The onset of a primary eclipse event shortly after the start of the
observations brought the observed brightness down by about a magnitude.
This made it evident that \object{HS~0705+6700} must be an eclipsing binary.
This was confirmed in the following night when both primary and secondary
minima were monitored.

\section{Observations}
\label{obs}

\subsection{Photometry}
\label{obsphot}

The initial observations were made on 5 and 6-Oct-2000 at the Nordic Optical
Telescope (NOT) with the Andalucia Faint Object Spectrograph and Camera
(ALFOSC), equipped with a Loral, Lesser thinned, 2048$\times$2048 CCD chip,
and modified with our own control software to be able to observe in high-speed
multi-windowing mode. The sky area available for locating a reference star is
limited to the area of the chip: $\sim$ 6.5$\times$6.5 arcminutes$^{2}$.
On 5-Oct-2000, the observations started at UT 05:38 and covered a time 
period of 3590\,s, and on 6-Oct-2000 measurements started at UT 03:28 and 
extended over 10280\,s.

We observed \object{HS~0705+6700} using three reference stars for constructing
the relative light curve. The observations were made with a Bessel $B$-band
filter in order to optimize for detection of low level pulsations. A cycle
time of 20\,s was used, allowing an actual integration time of 16.5\,s.
A total of 634 exposures were collected in the nights of 5 and 6-Oct-2000.
The data were reduced online using the Real Time Photometry (RTP) program
developed by one of us (RO) as part of his PhD project (\O stensen~ 
\cite{rothesis}). More details about this software are given elsewhere
(\O stensen et al.~\cite{ost01a}).

The $R$-band photometry of \object{HS~0705+6700} was obtained on 2-Nov-2000
at the Calar Alto Observatory using the 2.2m telescope with the CAFOS 
instrument. Observations started at UT 01:57, and the total length of the data 
set was 12841\,s. A total of 598 measurements were made with a cycle time of
21\,s (the integration time was 10\,s). Each exposure, taken through a Johnson 
$R$ filter, is a 2\,x\,2 binned subframe of the 2k\,x\,2k blue-sensitive,
back-illuminated SITe chip. The field around the target contains several other 
stars, five of which were used as reference stars. The basic data reduction 
as well as the relative aperture photometry was performed using the IDL 
software TRIPP (Schuh et al.~\cite{schuh99}). The 1\,$\sigma$ error of the 
resulting normalized differential light curve, derived from the variations 
of all relative reference light curves, is 0.013. 

Following the $B$ and $R$ light curves obtained in October and 
November 2000, additional photometric observations were collected with 
various smaller telescopes (on Tenerife, in Norway, Poland, and Greece) 
in five different nights scattered over the next five months (November 2000
to March 2001). Nine more primary minima were covered. The measurements 
were made with CCD cameras in different filters ($R, I$), and in one case 
with a classical photo-multiplier tube (PMT) photometer in unfiltered light.
Dates and other details of this complementary photometry are contained in 
Table~\ref{minimum}.

\begin{table}[h]
\begin{center}
\caption{Spectroscopic observations (heliocentric Julian Date for
         mid-exposure, length of exposure, and heliocentric radial velocity.}
\label{rv_tab}
\begin{tabular}{ccr}
\hline\noalign{\smallskip}
 hel.JD (mid) & exposure &   hel. RV       \\
 -2451900     & time (s) &   (km s$^{-1}$) \\[1mm]
\hline\noalign{\smallskip}
79.55107      &  900     &      56.5 \\
79.56562      &  900     &     -18.1 \\
79.57960      &  900     &     -74.5 \\
79.59264      &  900     &    -110.1 \\
80.46672      &  600     &    -114.5 \\
81.31732      &  600     &    -120.2 \\
81.32809      &  600     &    -106.4 \\
81.34120      &  900     &     -38.2 \\
81.35216      &  600     &       4.6 \\
81.36139      &  600     &      52.5 \\
81.37098      &  600     &      32.9 \\
81.38026      &  600     &       4.9 \\
81.39023      &  600     &     -45.7 \\
81.39943      &  600     &     -81.6 \\
81.40900      &  600     &    -116.6 \\[1mm]
\hline
\end{tabular}
\end{center}
\end{table}

\subsection{Spectroscopy}
\label{obsspec}

\begin{table*}[t]
\begin{center}
\caption{Times of primary minima derived from CCD and PMT photometric
         observations}
\label{minimum}
\begin{tabular}{cllrlcclll}
\hline\noalign{\smallskip}
Date & HJD & Error & Epoch & $O-C^{\mathrm{a}}$ & Method & Filter &
       Type & Telescope$^{\mathrm{b}}$ & Observers$^{\mathrm{c}}$ \\[1mm]
\hline\noalign{\smallskip}
05-10-00 & 2451822.75978 & 0.00005 &    0 & $-$0.00004 & Kwee     & B & CCD 
                                                        & NOT 2.5m & R\O, JES \\
06-10-00 & 2451823.71648 & 0.00010 &   10 & +0.00019 & Kwee     & B & CCD 
                                                        & NOT 2.5m & R\O, JES \\
02-11-00 & 2451850.59301 & 0.00011 &  291 & +0.00001 & parabola & R & CCD 
                                                        & CA 2.2m  & SS, JD  \\
02-11-00 & 2451850.68866 & 0.00012 &  292 & +0.00002 & parabola & R & CCD 
                                                        & CA 2.2m  & SS, JD  \\
30-11-00 & 2451878.71316 & 0.00003 &  585 & +0.00005 & Kwee     & - & PMT  
                                                        & IAC 0.8m & FJ      \\
09-01-01 & 2451919.2670  & 0.0002  & 1009 & $-$0.00029 & Kwee     & I & CCD 
                                                        & Skb 0.6m & SZ      \\
09-01-01 & 2451919.3627  & 0.0010  & 1010 & $-$0.00024 & parabola & I & CCD 
                                                        & Skb 0.6m & SZ      \\
09-01-01 & 2451919.4586  & 0.0001  & 1011 & +0.00002 & Kwee     & I & CCD 
                                                        & Skb 0.6m & SZ      \\
15-02-01 & 2451956.4738  & 0.0013  & 1398 & $-$0.00004 & parabola & R & CCD 
                                                        & Crc 0.5m & SZ      \\
16-02-01 & 2451956.5697  & 0.0011  & 1399 & +0.00022 & parabola & R & CCD 
                                                        & Crc 0.5m & SZ      \\
17-02-01 & 2451957.5274  & 0.0002  & 1409 & +0.00145 & Kwee     & R & CCD 
                                                        & Crc 0.5m & SZ      \\
27-03-01 & 2451996.3586  & 0.0001  & 1815 & +0.00011 & Kwee     & R & CCD 
                                                        & Kry 1.2m & SZ      \\
28-03-01 & 2451997.3150  & 0.0001  & 1825 & +0.00004 & Kwee     & R & CCD 
                                                        & Kry 1.2m & SZ \\[1mm]
\hline
\end{tabular}
\end{center}
\begin{list}{}{}
\item[$^{\mathrm{a}}$] Relative to ephemeris~(1)
\item[$^{\mathrm{b}}$]
   NOT: Nordic 2.5m, La Palma; CA: Calar Alto 2.2m, Spain;
   IAC: IAC 0.8m, Teide, Tenerife; Skb: Skibotn 0.6m, Norway;
   Crc: Cracow Jagiellonian Obs. 0.5m, Poland; Kry: Kryonerion 1.2m, Greece
\item[$^{\mathrm{c}}$]
   R\O: Roy \O stensen; JES: Jan-Erik Solheim; SS: Sonja Schuh; JD: Jochen 
   Deetjen; FJ: Frank Johannessen; SZ: Stanislaw Zola
\end{list}
\end{table*}

Optical spectra of \object{HS~0705+6700} were obtained in March 2001 at the
DSAZ observatory at the 3.5m telescope with the Twin spectrograph in dichroic
mode covering the wavelength ranges 3925\AA\ to 5015\AA\ in the blue channel
and 5975\AA\ to 7070\AA\ in the red channel at a spectral resolution of 1\AA. 
Details of the observations are given in Table~\ref{rv_tab}.
The data were reduced with the MIDAS package distributed by the European 
Southern Observatory ESO.

\section{Photometric analysis}
\label{anaphot}

\subsection{Ephemeris}
\label{ephem}

\begin{figure*}
\centering
\includegraphics[width=17cm]{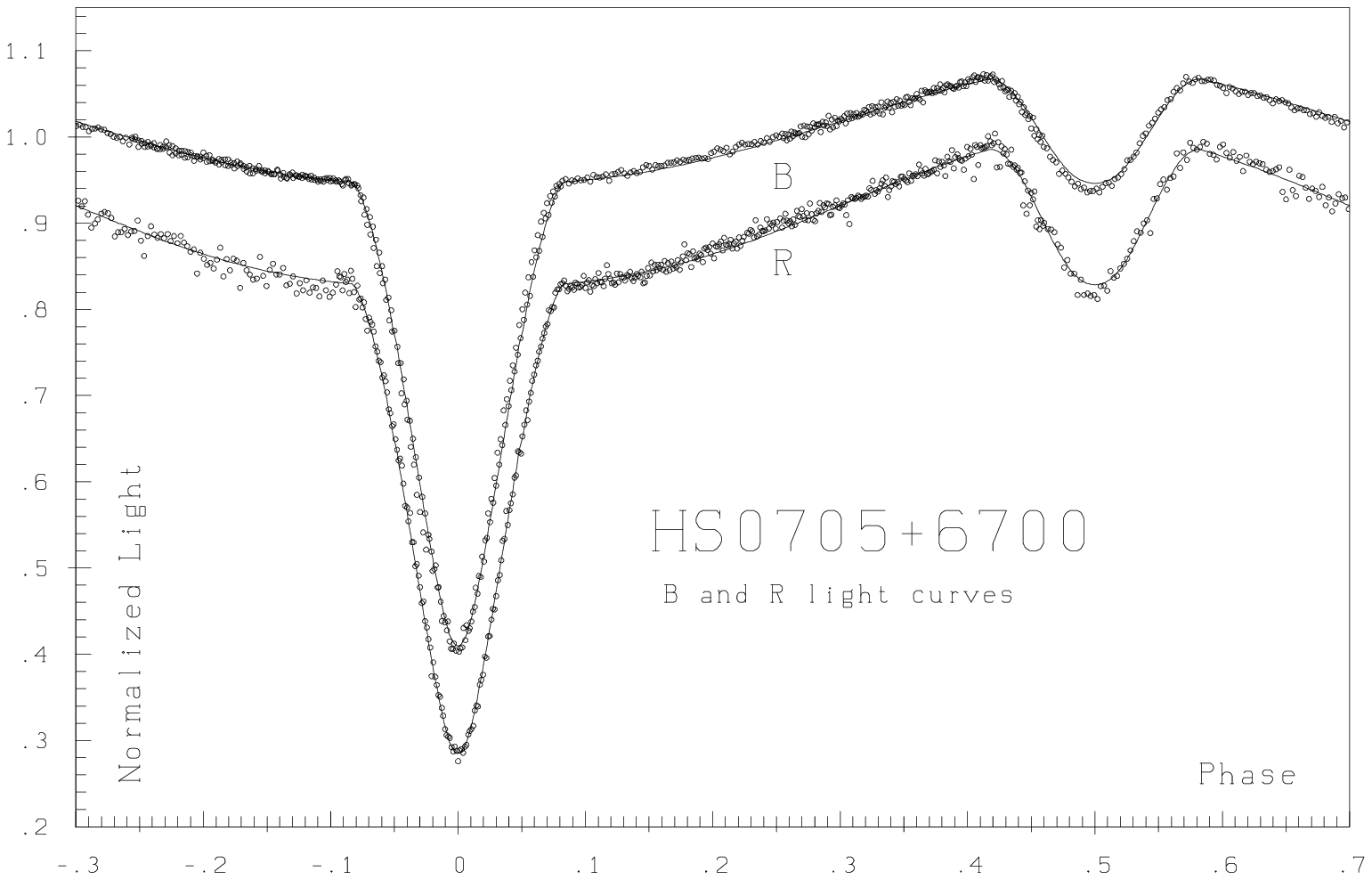}
\caption[]{$B$ and $R$ light curves of \object{HS~0705+6700} obtained on
           Oct. 5/6 ($B$) and Nov. 2, 2000 ($R$); the solid lines represent
           the best fit simultaneous solution according to the parameters
           listed in Table~\ref{lcparams}; the $R$ light curve is offset by 
           0.1 intensity units relative to the $B$ curve.}
\label{lc}
\end{figure*}

Our $B$ and $R$ CCD photometry of October-November 2000 provided clear
evidence that \object{HS~0705+6700} is a short-period eclipsing binary.
The complete phase coverage and high time resolution of the CCD light curves
allowed for an accurate determination of minimum times and a period analysis.

The CCD $B$ measurements carried out with the 2.5m Nordic Optical Telescope
covered a complete primary eclipse in each of the two nights of October 5 
and 6, 2000. The CCD $R$ photometry obtained with the CAFOS instrument 
attached to the Calar Alto 2.2\,m telescope on November 2, 2000, continuously
extended over more than one complete orbital cycle and covered two subsequent
primary minima. Already from these four minimum times we derived an orbital
period of 8263.88 seconds.

For a more refined period analysis nine more primary minima were used, 
which had been observed with several smaller telescopes between November 
2000 and March 2001 (see Table~\ref{minimum}). Including these minima
means a considerable broadening of the time base, allowing for more 
accurate period determination. The minimum times were determined either by
the Kwee-van~Woerden method (\cite{kwee56}) or by fitting parabolas to the
cores of the minima. Table~\ref{minimum} lists the minimum times together
with other observational details.

The standard deviation from a linear $O-C$ representation amounts to
$1.5\cdot10^{-4}$ days ($\approx$ 13 sec), which is less than the mean time 
resolution between subsequent CCD exposures when monitoring the light 
curves (mostly about 20 sec). Eight minima lie within a 1~$\sigma$ belt, and
four others within a 2~$\sigma$ belt. There is only one primary minimum (the
one of February 17, 2001, at epoch 1409) deviating by about 10~$\sigma$ from
the linear regression. This timing was therefore not used for the period
determination. From the remaining 12 primary minima distributed over a time
span of nearly 6 months, the following linear elements for the heliocentric
primary minimum were derived:
\begin{center}
HJD $2451822\fd75982 + 0\fd09564665 \cdot E$ \\
\hspace{1.9cm} $\pm 22$ \hspace{1.3cm} $\pm 39$
\end{center}

This ephemeris was applied for the phasing of the light curves and of the
spectroscopic observations analyzed in this paper.

\subsection{Light curve solution}
\label{lcsol}

The $B$ and $R$ light curves obtained with the 2.5m NOT and 2.2m Calar Alto 
telescopes, respectively, were analyzed with the MORO code (Drechsel et al.~
\cite{drechsel95}). This light curve solution program uses the Wilson-Devinney 
logistical approach (\cite{wildev71}), but is based on a modified 
Roche model for the binary structure, taking into account the radiative 
interaction between the components of hot, close binaries. Also, the 
differential corrections procedure for parameter optimization is replaced 
by the more powerful {\it simplex} algorithm, which was first applied to
eclipse light curve solutions by Kallrath \& Linnell (\cite{kallrath87}). For
more details and application examples see, e.g., Drechsel (\cite{drechsel00})
or Drechsel et al.~(\cite{drechsel95}).

The $B$ and $R$ light curves were represented by normal points formed by 
binning the fluxes of individual measurements over narrow time intervals.
One reason for this was to reduce the observational scattering to a 
smoother run, and on the other hand the computational performance was 
improved. The phase resolution of the normal points was generally chosen as 
0.005 phase units (41 sec), except for the core of the primary minimum, 
where a step width of 0.002 phase units (16 sec) was used. The total binning 
window size was 0.02 phase units (165 sec). Hence the 634 and 598 
individual measurements of the $B$ and $R$ curves, respectively, were 
reduced to 212 normal points in each passband. Typically, 12 individual 
observations were averaged to form a single normal point. The input light 
curves (in intensity units) were normalized to unity at quadrature phase
0.25.

We used the principal Wilson-Devinney mode 2, which poses no restrictions 
on the presumed system configuration (detached or semi-detached), 
and links the luminosity of the secondary component to its effective 
temperature by means of the Planck law. The latter aspect is of nearly no 
importance here, because the luminosity ratio between primary and secondary 
flux contributions is so huge in this system (L$_1$/L$_2$ is of order 
10$^4$ for B and 10$^3$ for R passbands), that the secondary is 
photometrically only evident through the reflection effect and shadowing of 
primary light. The $B$ and $R$ light curves were solved simultaneously to 
yield consistent solution parameters.

Since the total number of light curve parameters is considerably large
$(12 + 5 \cdot n)$, where $n$ is the number of spectral passbands (filter 
light curves), i.e. equals 22 for our case of a simultaneous $B$ and $R$ fit, 
it is important to reduce the free parameter set by consideration of 
spectroscopic and theoretical boundary conditions and other consistency 
checks. Otherwise, too many free parameters would easily tend to produce
underdetermined solutions with no guarantee of uniqueness.

As was obvious from the sinusoidal shape of the radial velocity curve
(Sect.~\ref{rvcurve}) and from the position of the secondary minimum at 
exactly phase 0.50, it was implicitly assumed that we are dealing with
circular orbits $(e = 0)$ of synchronously rotating components in this very 
close binary system, for which an extremely short synchronization time scale
of a few decades is expected (Zahn \cite{zahn77}). According to the early
spectral type of the primary we assumed a bolometric albedo $A_1 = 1$; also
its gravity darkening exponent was fixed at $g_1 = 1$ as expected for
radiative outer envelopes (von Zeipel \cite{zeipel24}). For the cool convective
secondary, $g_2$ was set to 0.32 according to Lucy (\cite{lucy67}). Already
after the first trial runs with the secondary albedo $A_2$ as a free
parameter, it became obvious that the enormous reflection effect evident as
a broad orbital hump centered on secondary eclipse, with an amplitude of
about 12\,\% of maximum light, requires a ``mirror-like'' surface of the
tidally locked secondary in the heated area facing the primary. Reasonable
representations were only possible for complete reradiation (energy
conservation), i.e. $A_2$ was fixed at 1.0 in all further runs. Linear limb
darkening coefficients $x_1(B,R)$ for the primary were interpolated from the
tables of Wade \& Rucinski (\cite{wade85}) and fixed at $x_1(B) = 0.26$ and
$x_1(R) = 0.19$. Since it was known from an analysis of similar sdOB
binaries with strong reflection effect (Hilditch et al.~\cite{hilditch96})
that the limb darkening coefficients of cool stars irradiated 
by a hot companion can strongly deviate from ``normal'' values of very cool 
dwarf stars, it was decided to treat $x_2$ as an adjustable parameter. From 
our spectroscopic analysis (Sect.~\ref{specanal}) it was clear that the sdB 
primary must have an effective temperature close to 30\,000~K. Though $T_1$ 
was always used as an adjustable quantity, it actually did not vary much 
during numerous optimization runs, and mostly remained within a narrow 
range between about 29\,000 and 30\,000~K, compatible with the spectroscopic 
error margin.

   The remaining set of adjustable parameters comprises the orbital 
inclination $i$, the mass ratio $q = M_2/M_1$, the Roche potentials
$\Omega_1$ and $\Omega_2$ of the two stellar surfaces, the color-dependent 
luminosity $L_1$ of the primary and of a possible third light contribution 
$l_3$, the effective temperatures $T_1$ and $T_2$, and the radiation 
pressure parameter $\delta_1$ (see Drechsel et al.~\cite{drechsel95});
$\delta_2$ was set equal to zero because the secondary temperature is low
(about 3\,000~K), which means that radiation pressure forces exerted by 
this star are negligible. Finally, $L_2$ was not adjusted as an independent 
parameter, but recomputed from the secondary radius $r_2$ and its effective 
temperature.

\begin{figure}
\centering
\includegraphics[width=8.8cm]{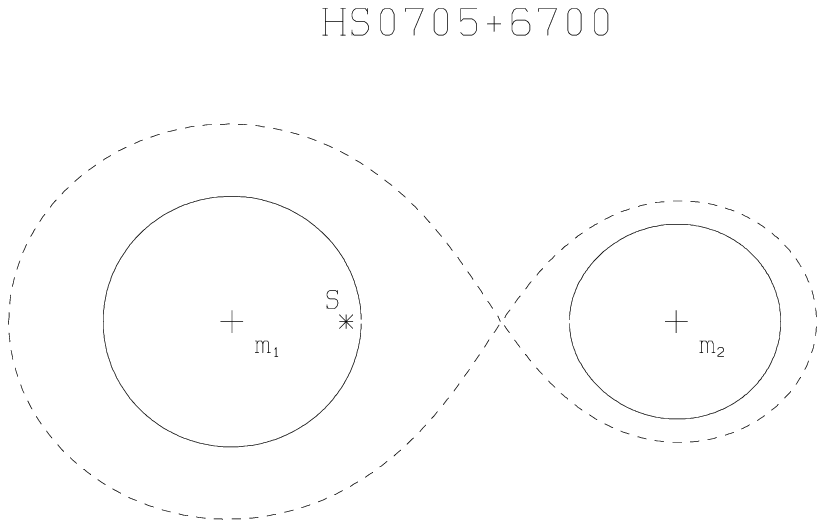}
\caption[]{Meridional intersection of the equipotential structure of
           \object{HS~0705+6700} corresponding to the best fit photometric
           solution for a mass ratio $q = 0.278$; the inner critical Roche
           lobe is shown as dashed line.}
\label{meri}
\end{figure}

   A convergent simultaneous solution of the $B$ and $R$ light curves was 
achieved after several trial runs using different start parameter sets, 
most of which lead to an essentially identical final solution.
Table~\ref{lcparams} gives a list of the fixed and adjusted parameters. The 
corresponding fitted light curves together with observations are shown in 
Fig.~\ref{lc}. It is obvious from the graph that the overall run of the $B$ 
and $R$ light curves is very well reproduced by the best fit theoretical 
curves (solid lines). Especially the shape of the primary eclipse minima
and the pronounced reflection effect are fitted very well. A minor 
systematic deviation is apparent in the secondary minima: the fitted curves 
are somewhat too shallow in the center of the minimum, though the effect is 
hardly larger than the observational error margins. A physical explanation may 
be found in the treatment of the reflection effect. Our Wilson-Devinney-based
solution code uses bolometric albedos and black body radiation to characterize
the reflectivity of the irradiated surface layers. The upper limit of albedo
is therefore restricted to 100\,\% for complete reradiation of the incident
flux, and in our case the adjusted albedo of the cool secondary was actually
limited by this constraint. Albedos larger than 1 would be ``unphysical'', 
if local non-Planckian deviations of the intensity distribution are not 
accounted for. Removing the constraint on $A_2$ would result in an 
increased adjusted value of $A_2$ of $\sim 1.1$, and a deepening of the 
secondary minimum by a few millimag. Because of the tininess of this effect 
and due to the  highly complex physical problems connected with the external
irradiation of stellar atmospheres we preferred not to use such purely
phenomenological approximation and remained with the classical reflection 
effect approach. Also, a suspected shift of the secondary minimum in the 
$B$ curve by a few $10^{-3}$ phase units relative to phase 0.50 cannot be 
modeled without introducing additional free parameters. If real, it would 
conceivably be caused by a non-axisymmetrical brightness distribution of the
heated surface area of the secondary component.

The $1~\sigma$ standard deviation of the normal points from the computed light
curve amounts to about 0.004~mag. It should be noted that the best fit in the
$B$ passband ($\lambda_{\mathrm{eff}}$ = 435~nm) requires a third light
contribution of 3.7\,\%, while the $R$ curve
($\lambda_{\mathrm{eff}}$ = 641~nm) is fully consistent with $l_3 = 0$.
The same effect was observed in the closely related sdB binary \object{HW~Vir}
(Wood et al.~\cite{wood93}), in which the $U$ and $B$ filter measurements
deviate from the computed curves resulting from a simultaneous $UBVR$ fit by a
similar amount, whereas the $R$ curve shows no such systematic effect.
A possible explanation might arise from the strong reflection effect in
both systems: though energy is conserved in the bolometric reradiation of the
heated surface layers, there could be some frequency redistribution of photons
inherent in the complex scattering process in a way that higher-energy photons
have larger scattering cross sections than those incident with longer
wavelengths and/or are transformed into lower-energy photons by various
absorption and reemission processes.

\begin{table}
\begin{center}
\caption{Light curve solution of \object{HS~0705+6700}}
\label{lcparams}
\begin{tabular}{ll}
\hline\noalign{\smallskip}
\multicolumn{2}{l}{Fixed parameters:}       \\[1mm]
\hline\noalign{\smallskip}
$A_1\ ^\mathrm{a}$                 & 1.0    \\
$A_2\ ^\mathrm{a}$                 & 1.0    \\
$g_1\ ^\mathrm{b}$                 & 1.0    \\
$g_2\ ^\mathrm{b}$                 & 0.32   \\
$x_1(\mathrm B)\ ^\mathrm{c}$      & 0.26   \\
$x_1(\mathrm R)\ ^\mathrm{c}$      & 0.19   \\
$\delta_2\ ^\mathrm{d}$  & 0.0    \\[1mm]
\hline\noalign{\smallskip}
\multicolumn{2}{l}{Adjusted parameters:}    \\[1mm]
\hline\noalign{\smallskip}
$i$                      & $84\fdg4    \pm 0\fdg3  $ \\
$q (= M_2/M_1)$          &  $0.278     \pm 0.019   $ \\
$\Omega_1$               &  $3.764     \pm 0.066   $ \\
$\Omega_2$               &  $2.606     \pm 0.051   $ \\
$T_\mathrm{eff}(1)$      &  29\,600\,K$\pm 800$\,K   \\
$T_\mathrm{eff}(2)$      &   2\,900\,K$\pm 600$\,K   \\
$L_1$(B) $^\mathrm{e}$   &  $0.99997   \pm 0.00003 $ \\
$l_3$(B) $^\mathrm{f}$   &  $3.7\,\%   \pm 0.6\,\% $ \\[1mm]
$L_1$(R) $^\mathrm{e}$   &  $0.99969   \pm 0.00031 $ \\
$l_3$(R) $^\mathrm{f}$   &  $0.2\,\%   \pm 0.2\,\% $ \\[1mm]
$x_2$(B)                 &  $0.42      \pm 0.12    $ \\
$x_2$(R)                 &  $0.62      \pm 0.17    $ \\
$\delta_1\ ^\mathrm{d}$  &  $0.025     \pm 0.011   $ \\[1mm]
\hline\noalign{\smallskip}
\multicolumn{2}{l}{Roche radii: $^\mathrm{g}$}\\[1mm]
\hline\noalign{\smallskip}
$r_1(\mathrm{pole})$     & $0.279 \pm 0.004 $ \\
$r_1(\mathrm{point})$    & $0.287 \pm 0.002 $ \\
$r_1(\mathrm{side})$     & $0.283 \pm 0.003 $ \\
$r_1(\mathrm{back})$     & $0.285 \pm 0.003 $ \\[1mm]
$r_2(\mathrm{pole})$     & $0.219 \pm 0.001 $ \\
$r_2(\mathrm{point})$    & $0.238 \pm 0.004 $ \\
$r_2(\mathrm{side})$     & $0.225 \pm 0.002 $ \\
$r_2(\mathrm{back})$     & $0.239 \pm 0.005 $ \\[1mm]
\hline
\end{tabular}
\end{center}
\begin{list}{}{}
\item[$^\mathrm{a}$] bolometric albedo
\item[$^\mathrm{b}$] gravitational darkening exponent
\item[$^\mathrm{c}$] linear limb darkening coefficient; theoretical
                     value taken from Wade \& Rucinski(\cite{wade85})
\item[$^\mathrm{d}$] radiation pressure parameter,
                     see Drechsel et al.~\cite{drechsel95}
\item[$^\mathrm{e}$] relative luminosity $L_1/(L_1 + L_2)$; $L_2$ is not 
                     independently adjusted, but recomputed from $r_2$ and 
                     $T_\mathrm{eff}(2)$
\item[$^\mathrm{f}$] fraction of third light at maximum
\item[$^\mathrm{g}$] fractional Roche radii in units of separation of
                     mass centers
\end{list}
\end{table}

\begin{figure*}
\centering
\includegraphics[width=15cm]{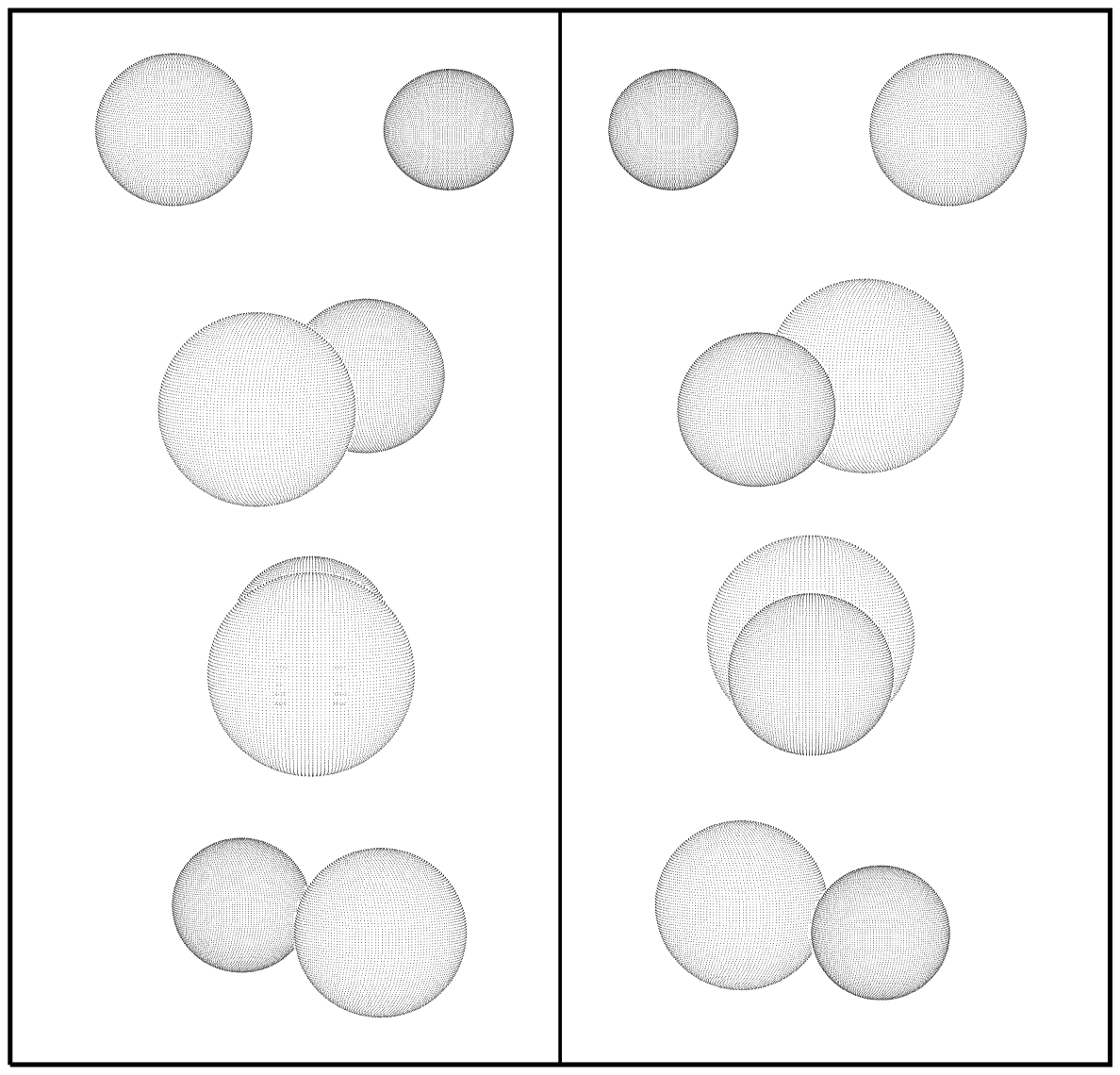}
\caption{Aspects of \object{HS~0705+6700} at orbital phases 0.25, 0.45, 0.50,
         0.575 (left column, from top to bottom), 0.75, 0.95, 0.00, 0.075
         (right column), viewed under the orbital inclination angle of
         84$\fdg4$.}
\label{movie}
\end{figure*}

The photometric solution yields a close detached system configuration 
for \object{HS~0705+6700}. A meridional intersection of the surface
equipotentials and the inner critical Roche lobe is shown in Fig.~\ref{meri}.
The cool secondary is found to be relatively closer to its Roche lobe than the 
primary, which appears only slightly distorted by tidal interaction
($r_{\mathrm{point}}/r_{\mathrm{pole}} \approx$ 1.03 and 1.09, respectively,
for the primary and secondary components). 3D-simulated aspects of the system
at various orbital phases under a viewing angle corresponding to the derived 
orbital inclination of 84$\fdg4$ are displayed in Fig.~\ref{movie}. 

Though the {\it simplex} algorithm is a more powerful and numerically stable 
parameter optimization method than the differential corrections procedure, 
which is usually used in combination with the Wilson-Devinney approach, 
it also cannot completely rule out the possibility of convergence into a 
local minimum in the multi-dimensional parameter space. For this reason we 
experimented with a multitude of different start parameter sets to verify 
that our final solution corresponds to the deepest global minimum. We found 
good evidence that this solution indeed turned out to be preferred in most 
of our trial runs. We should, however, mention that there were a few other 
solutions which differ from our best fit mainly in their values 
of the mass ratio $q$. While the best fit gave $q = 0.278$,  other solutions
with larger alternative values of $q$ = 0.346, 0.460, and 0.594 were also
found. Yet the standard deviations of the fits for $q = 0.46$ and 0.59 are
worse by about 5-10\,\%, and they are also not consistent with our
spectroscopic results (inconsistent value of the mass function, see
Sect.~\ref{specanal}). The photometric solution for $q = 0.35$ is of nearly
identical numerical quality like our preferred one for $q = 0.28$, but in
order to match the observed value of the mass function of $f(m) = 0.0063$\Msun,
the primary mass had to be assumed as $\leq 0.3$\Msun. Hence a value of
$M_2 \leq 0.1$\Msun\ would result. According to low-mass star models (e.g.,
Dorman et al.~\cite{dorman89}), the mass-radius relation would then yield a
secondary radius of the order of $0.10-0.12$\Rsun, which would be incompatible
with the photometrically derived value.

It is a common problem inherent to light curve solution techniques that it 
is not easy to identify the deepest global minimum in the multi-dimensional 
parameter space. Since several parameters used to model the light curves 
are highly correlated, often a multitude of nearly equally good
solutions can be found for different solution parameter sets. In our case, 
we also end up with various solutions with a comparable goodness of fit, which 
mainly differ in their $q$ values (see above). The light curve of
\object{HS~0705+6700} is very similar to those of \object{HW~Vir} and
\object{PG~1336-018}. Wood et al.~(\cite{wood93}) and Kilkenny et
al.~(\cite{kilkenny98}) both encountered the same problem when solving the 
light curves of these systems and were able to obtain good solutions for a
broad range of mass ratios. The {\it simplex} parameter optimization 
algorithm incorporated in our MORO solution code is already a powerful tool 
with respect to the ability to automatically scan the multi-dimensional
space of adjustable parameters for the deepest minimum within broad 
parameter ranges. However, the dependency of the goodness of fit on the chosen 
parameter set can be highly complex, or the relation is locally
``flat-bottomed'', so that a unique solution does not exist. 

To demonstrate this behaviour a number of solutions with different mass 
ratios were enforced by fixing $q$ at various values in the surroundings of 
our best solution. The goodness of fit is measured by the standard 
deviation $\sigma_{\rm fit}$, which is defined by
\mathindent1.9cm
\begin{displaymath}
\sigma^2_{\rm fit} = \frac{n}{n-m} \frac{\sum_{i=1}^n w_i 
                                       (O_i-C_i)^2}{\sum_{i=1}^n w_i} ,
\end{displaymath}
where $n$ is the number of observations (normal points) and $m$ the number 
of adjustable quantities; $O_i-C_i$ are the residuals between normal points 
and fitted curve, and $w_i$ are the individual weights of the normal points.
The {\it simplex} routine uses these same $\sigma_{\rm fit}$ values also 
internally during each iteration step to control the optimization procedure.
Table~\ref{sigma} summarizes the final $\sigma_{\rm fit}$ values of
convergent solutions obtained for a grid of arbitrary $q$ values, which were
kept fixed during these trial runs, while the rest of light curve parameters
was adjusted to yield the best possible numerical representation of the
observed $B$ and $R$ curves. We give $\sigma_{\rm fit}$ values for the 
composite set of simultaneously adjusted $B$ and $R$ light curves,
$\sigma_{\rm tot}$, as well as the individual $\sigma_{\rm fit}$ values of 
the $B$ and $R$ curves, $\sigma_{\rm B}$ and $\sigma_{\rm R}$. The results 
for the trial solutions at $q$ = 0.20, 0.25, 0.30, 0.35, and 0.40 are 
compared with our selected best solution for $q$ = 0.278, which was 
obtained by adjusting $q$ together with the whole parameter set. It is 
evident that there are two ranges of $q$ around 0.28 and 0.35, where the
standard deviations occupy local minima in the parameter space. The trial
solution for $q$ = 0.35 (fixed) nearly coincides with one of our previously 
discussed unrestricted solutions with $q$ treated as a free parameter
(resulting in $q$ = 0.346). The distinction between the two alternative
$q$ values 0.28 and 0.35 apparently cannot be made solely based on numerical
standards, since the difference of the respective $\sigma$ values is too
small. Fortunately, in our case the available spectroscopic information (mass
function, mass-radius relation for the low-mass dwarf companion) could be used
to sort out inconsistent photometric solutions. The small error given in
Table~\ref{lcparams} for $q = 0.278 \pm 0.019$ simply reflects the formal fit
uncertainty within the local minimum in parameter space around $q = 0.28$,
and does not account for systematic effects like ambiguity of solutions.

\begin{table}
\begin{center}
\caption{Goodness of light curve fit for various fixed $q$ values compared 
         with $\sigma$ of best fit solution with adjusted $q$}
\label{sigma}
\begin{minipage}{6cm}
\begin{tabular}{lccc}
\hline\noalign{\smallskip}
 $q$ & $\sigma_{\rm B}$ & $\sigma_{\rm R}$ & $\sigma_{\rm tot}$ \\[1mm]
\hline\noalign{\smallskip}
0.20        & 0.00433 & 0.00429 & 0.00428 \\
0.25        & 0.00426 & 0.00426 & 0.00423 \\
{\bf 0.278} \footnote{adjusted $q$ of final best solution}
            & {\bf 0.00407} & {\bf 0.00403} & {\bf 0.00401} \\
0.30        & 0.00411 & 0.00414 & 0.00409 \\
0.35        & 0.00408 & 0.00402 & 0.00401 \\
0.40        & 0.00421 & 0.00398 & 0.00406 \\[1mm]
\hline
\end{tabular}
\end{minipage}
\end{center}
\end{table}

\section{Spectroscopic analysis}
\label{specanal}

The optical spectra allow to measure the radial velocity curve with good 
phase coverage as well as a quantitative spectral analysis to determine
the atmospheric parameters and the projected rotational velocity. 

\begin{figure*}
\centering
\includegraphics[width=17cm]{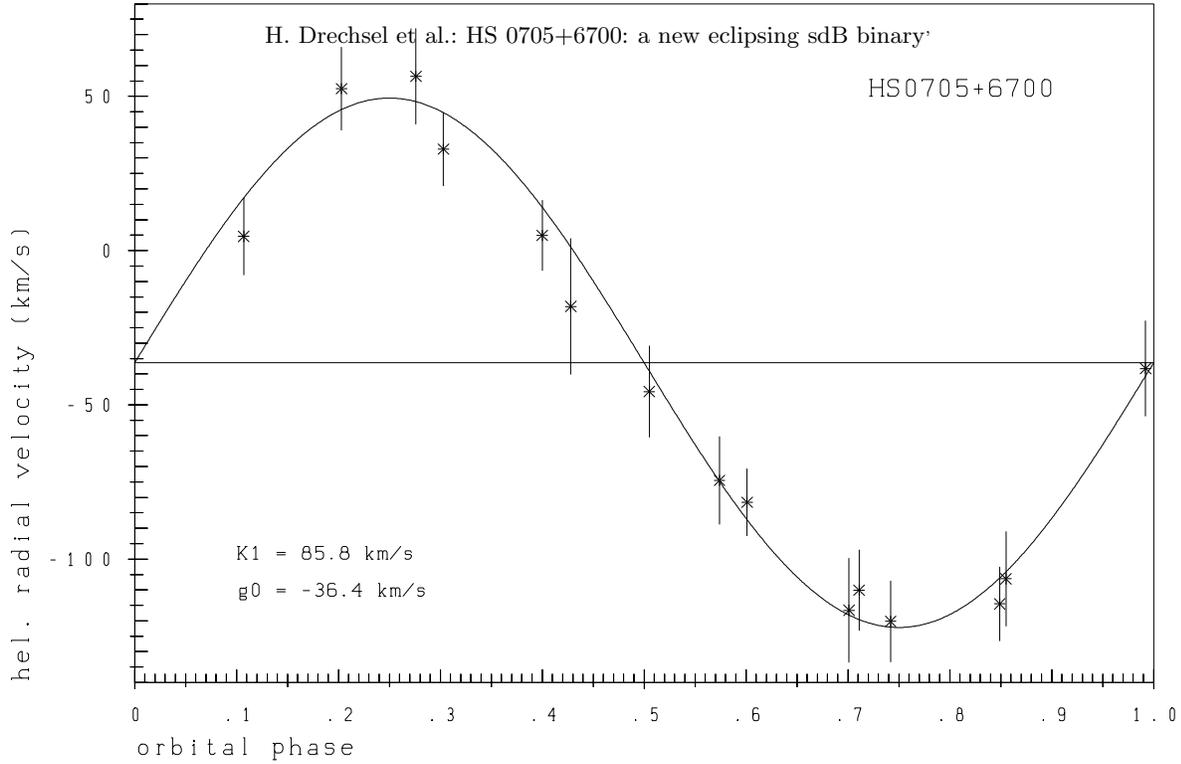}
\caption{Radial velocity curve of the sdB primary component in a circular 
         orbit; velocities are heliocentrically corrected; the semi-amplitude
         $K_1$ is 85.8\kms, the systemic velocity $\gamma_0 = -36.4$\kms;
         error bars correspond to $3~\sigma$ margins.}
\label{rv}
\end{figure*}

\subsection{Radial velocity curve}
\label{rvcurve}

Radial velocities from all Twin spectra were derived by cross-correlating
the observed helium and Balmer line spectra to a synthetic spectrum 
calculated from a model atmosphere (see below) and are listed in 
Table~\ref{rv_tab}. The measurements are accurate to about
$\pm 15$\kms\ (3\,$\sigma$).

Since this system is single-lined the analysis is straightforward. 
A comparison of the best fit radial velocity curve and the measured 
values is shown in Fig.~\ref{rv}. The semi-amplitude is 
$K_1 = 85.8 \pm 3.7$\kms and the systemic velocity
$\gamma_0 = -36.4 \pm 2.9$\kms.

From the period and semi-amplitude the mass function is derived: 
$f(m) = 0.00626 \pm 0.00081$\Msun.

\subsection{Spectrum fitting}
\label{specfit}

In order to improve the S/N ratio the Twin spectra were radial
velocity-corrected and then coadded. The coadded as well as the individual
spectra were analyzed to derive atmospheric parameters using NLTE model
atmospheres (Napiwotzki {\cite{napi97}) and a $\chi^2$ procedure described
by Napiwotzki et al.~(\cite{napi99}).

Matching the synthetic Balmer (H$\beta$ to H$\epsilon$) and He~I 
(4026\,\AA, 4471\,\AA\, and 4922\,\AA) line profiles to the observations 
resulted in a determination of effective temperature, gravity and He
abundance. The resulting fit for the coadded spectrum is displayed in
Fig.~\ref{teffg_plot} and gave T$_{\rm eff} = 28\,800 \pm 300$~K, 
$\log g = 5.40 \pm 0.04$ and $\log (n_{\rm He}/n_{\rm H}) = -2.68 \pm 0.05$.
The quoted errors are derived from the $\chi^2$ procedure applied to fit 
the line profiles. 

We also analyzed the individual spectra in the same way as described above.
Two spectra (at phases 0.701 and 0.855) were not useable for a quantitative
spectral analysis because of insufficient S/N. These two phases, however,
are covered by better spectra (at phases 0.711 and 0.849) anyway. The
results for the remaining 13 phases are summarized in Table~\ref{teffg},
and are plotted in Fig.~\ref{ph_t_g}. Because of the reflection effect the 
Balmer lines might be distorted by reflected light from the secondary.
Indeed, in the case of \object{HW~Vir} Wood \& Saffer (\cite{woosaf99}) found
that T$_{\rm eff}$ varied by 1500\,K and $\log g$ by 0.1~dex around the orbit.  
Two spectra of \object{HS~0705+6700} were taken near primary and another near
secondary minimum, for which we do not expect any contamination by 
reflected light.

The solution of the B light curve indicated the presence of third light.
This could cause a systematic error for the spectroscopic analysis.
The spectral characteristics of this light, however, are unknown. We made
an experiment on an individual spectrum and subtracted 3\% of the continuum 
from the blue spectra and repeated the fit. In this case the effective 
temperature is lowered by about the same amount as the 1$\sigma$ errors 
listed in Table~\ref{teffg}. The change in gravity is only a few hundredth of
a dex, less than the errors listed in this table. As possibly indicated by
Fig.~\ref{ph_t_g}, there might be a slight variation of T$_{\rm eff}$ with
orbital phase, but not so for $\log g$. Our spectra are probably of
insufficient quality to reveal any phase-dependent variations of the
atmospheric parameters at such low level.

We finally adopted T$_{\rm eff} = 28\,800 \pm 900$~K, $\log g = 5.40 \pm 0.1$
and $\log (n_{\rm He}/n_{\rm H}) = -2.68 \pm 0.15$ for the atmospheric
parameters of \object{HS~0705+6700}. The given errors were estimated from the
scatter of the $T_{\rm eff}$ and $\log g$ values derived from the individual 
spectra and hence would cover the bandwidth of any possibly present systematic
variation with phase.

We finally note that the spectroscopic $T_{\rm eff}$ is in excellent agreement
with the one following from the photometric solution (Sect.~\ref{lcsol}).

\begin{table}
\begin{center}
\caption{Atmospheric parameters of \object{HS~0705+6700}}
\label{teffg}
\begin{tabular}{clcc}
\hline\noalign{\smallskip}
phase   &  T$_{\rm eff}$ (K)  & $\log g$       & $-\log ({\rm He}/{\rm H})$ \\
\hline\noalign{\smallskip}
0.107   &  $27300 \pm  836$   & $5.41 \pm 0.10$  & $2.50 \pm 0.11$  \\
0.203   &  $27720 \pm  858$   & $5.29 \pm 0.11$  & $2.82 \pm 0.14$  \\
0.276   &  $29748 \pm  901$   & $5.48 \pm 0.14$  & $2.60 \pm 0.22$  \\ 
0.303   &  $29063 \pm  697$   & $5.30 \pm 0.10$  & $2.44 \pm 0.13$  \\
0.400   &  $28072 \pm  854$   & $5.28 \pm 0.11$  & $2.55 \pm 0.12$  \\
0.428   &  $28947 \pm 1600$   & $5.12 \pm 0.23$  & $2.99 \pm 0.49$  \\
0.505   &  $28414 \pm 1002$   & $5.30 \pm 0.13$  & $2.63 \pm 0.12$  \\
0.574   &  $29928 \pm  828$   & $5.43 \pm 0.14$  & $2.81 \pm 0.23$  \\
0.601   &  $28649 \pm  751$   & $5.59 \pm 0.11$  & $2.54 \pm 0.11$  \\
0.711   &  $30126 \pm  649$   & $5.51 \pm 0.11$  & $2.89 \pm 0.19$  \\
0.742   &  $28813 \pm  849$   & $5.50 \pm 0.12$  & $2.78 \pm 0.15$  \\
0.849   &  $28186 \pm  842$   & $5.26 \pm 0.10$  & $2.58 \pm 0.11$  \\
0.992   &  $27102 \pm 1009$   & $5.37 \pm 0.13$  & $2.94 \pm 0.20$  \\[1mm]
\hline\noalign{\smallskip}
Mean:   &  $28621 \pm  919$   & $5.37 \pm 0.12$  & $2.70 \pm 0.17$  \\[1mm]
\hline\noalign{\smallskip}
coadded &  $28755 \pm  280$   & $5.40 \pm 0.04$  & $2.68 \pm 0.05$  \\[1mm]
\hline
\end{tabular}
\end{center}
\end{table}

\begin{figure}
\includegraphics[width=8.8cm]{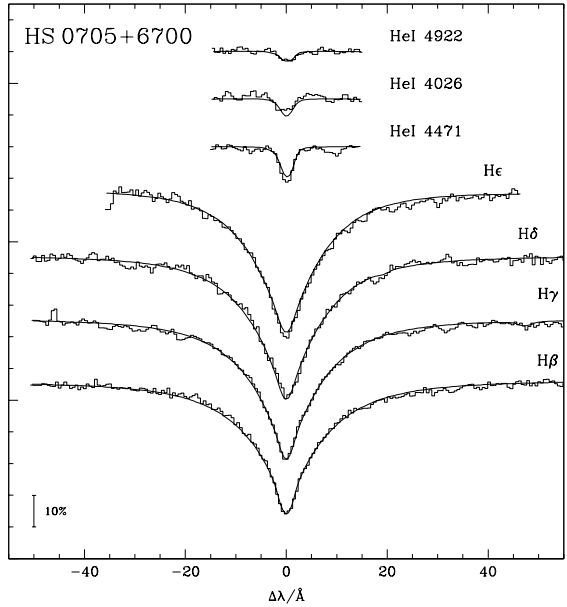}
\caption[]{Fit of the Balmer and helium lines in the coadded spectrum of 
           \object{HS~0705+6700} by synthetic spectra calculated from NLTE
           model atmospheres.}
\label{teffg_plot}
\end{figure}

\begin{figure}
\includegraphics[width=8.8cm]{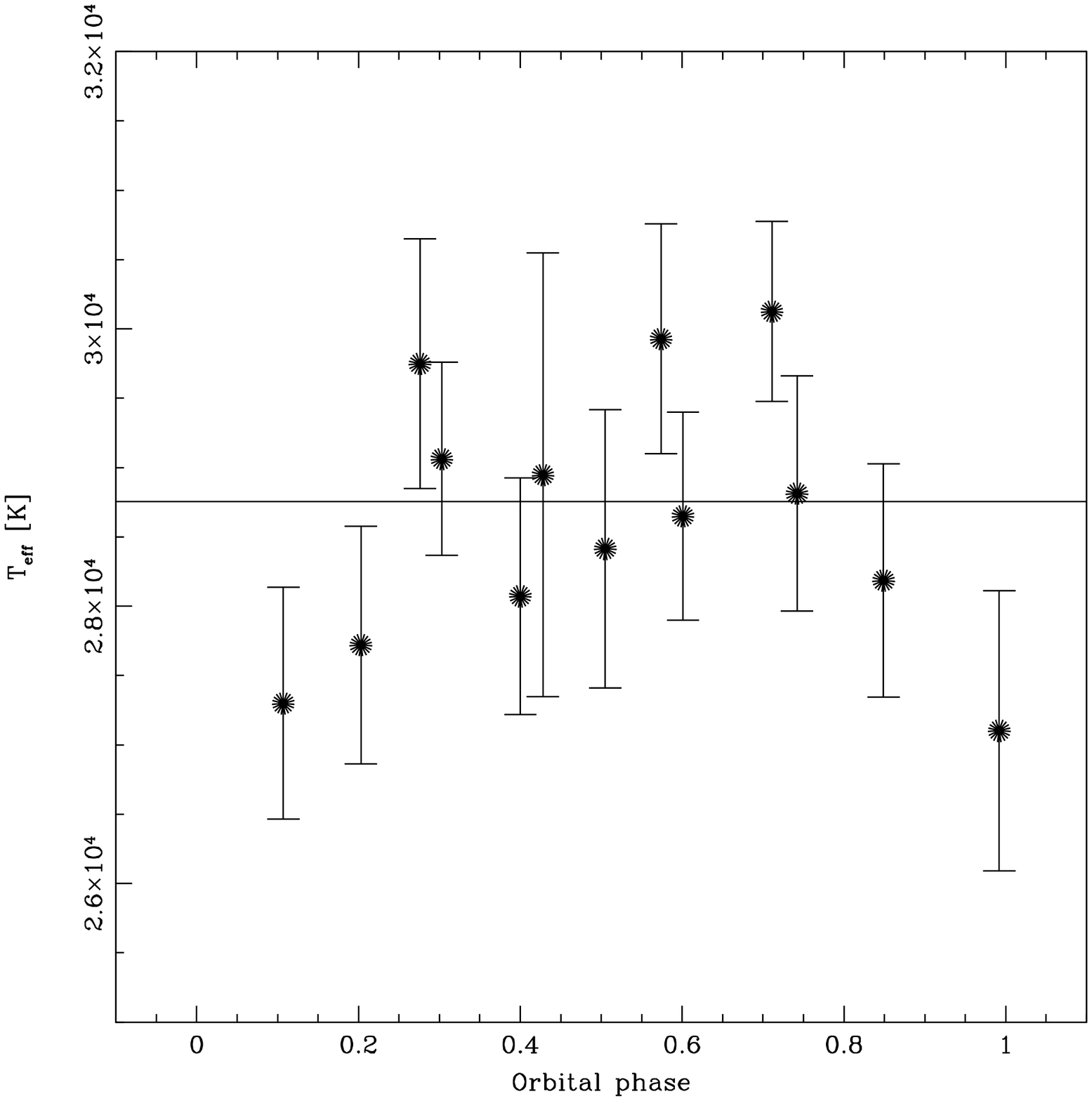}
\includegraphics[width=8.8cm]{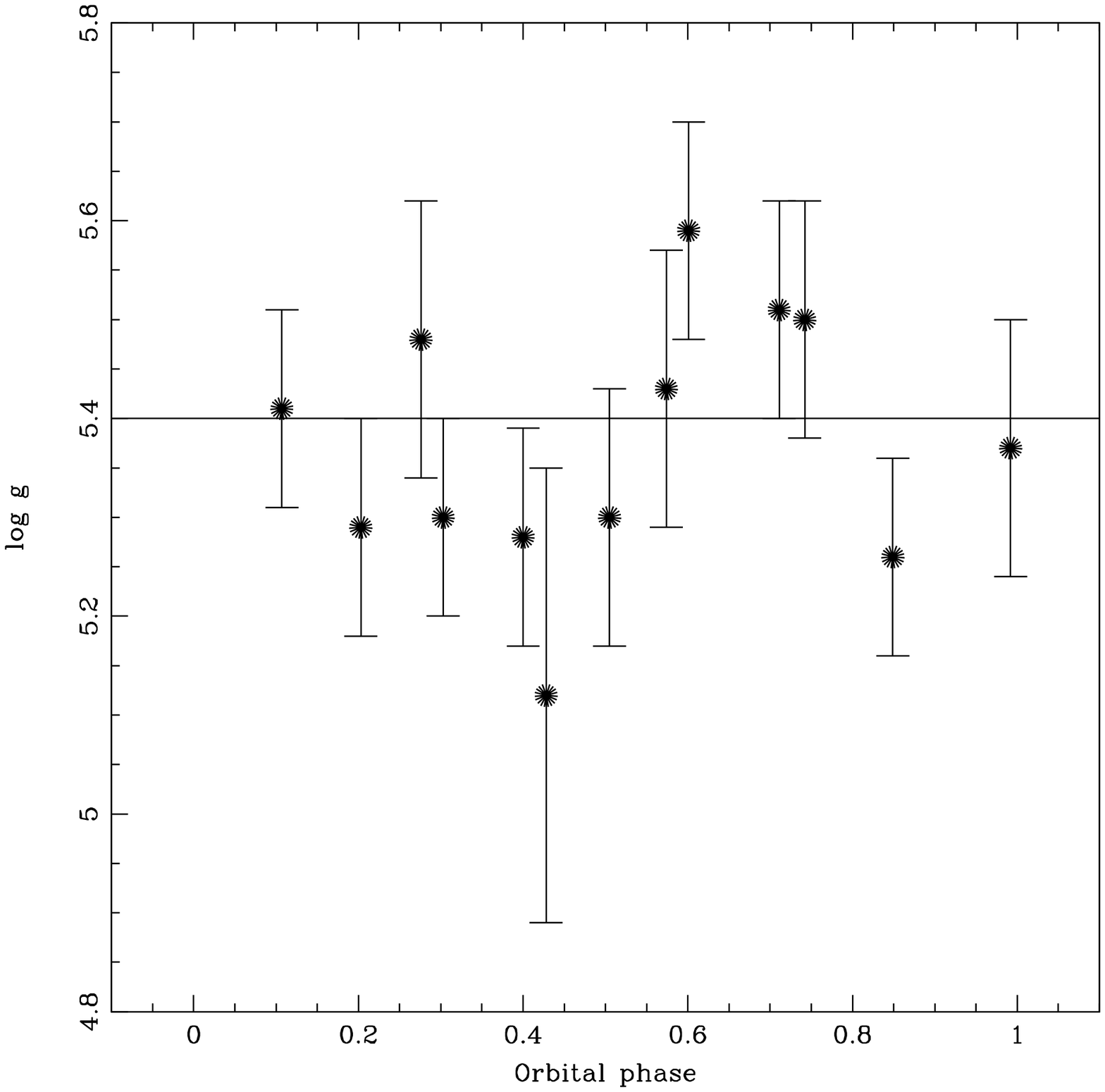}
\caption[]{Variation of atmospheric parameters with phase; the error bars 
           result from the $\chi^2$ fit procedure described by 
           Napiwotzki et al.~\cite{napi99}}
\label{ph_t_g}
\end{figure}

\subsection{Projected rotational velocity}

The line profiles of  H$\alpha$ and He~I\,6678\AA\ could be used to
determine the projected rotational velocity. We calculated the H$\alpha$ 
profiles from the final model and convoluted it with rotational profiles 
for various rotational broadening parameters. The best fit was again 
obtained by a $\chi^2$ procedure as described by Heber et al.~
(\cite{heber97}), which resulted in a projected rotational velocity of 
$v \sin i = 110^{+16}_{-13}$\kms\ (3\,$\sigma$ errors).
The corresponding fit is displayed in Fig.~\ref{halpha}. The rotation of
\object{HS~0705+6700} is expected to be tidally bound. In this case we can
estimate the projected rotational velocity from the orbital period and the
radius (see Sect. \ref{params}) to be $121^{+15}_{-13}$\kms\ in good
agreement with the result from spectrum synthesis. 

\begin{figure}
\includegraphics[width=8.8cm]{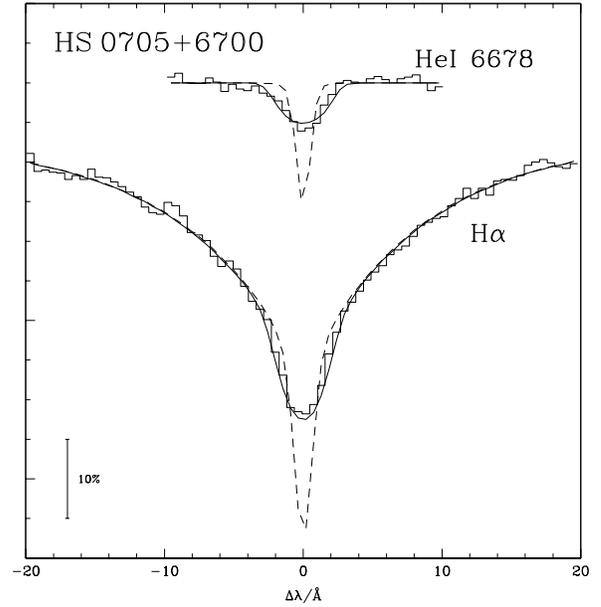}
\caption[]{Determination of the projected rotational velocity; synthetic 
           H$\alpha$ and He~I line profiles calculated from the finally
           adopted model and the best fit $v\,\sin i = 110$\kms\ are compared
           to the coadded spectrum; the same line profiles predicted for a
           non-rotating star are also shown as dashed lines.}
\label{halpha}
\end{figure}

\section{System parameters}
\label{params}

\subsection{HS~0705+6700}
\label{hs0705_params}

Even though the radial velocity amplitude of the companion is unknown we have
sufficient information to calculate absolute system parameters due to the 
$M-R$ relation provided by the $\log g$ determination (Sect.~\ref{specfit}).

From the light curve we have derived a period of 8263.87\,s, an inclination
angle of $i = 84\fdg4$, and a mass ratio $q = 0.278$. Furthermore, we determined
the radii for the binary components $R_1/a = 0.284$ and $R_2/a = 0.230$,
where $a$ is the separation of their mass centers. Using our measured
$K_1 = 85.8$\kms, $\log g_1 = 5.40$, and the inclination $i$, we derive
$M_1 = 0.483$\Msun, $M_2 = 0.134$\Msun, $R_1 = 0.230$\Rsun, $R_2 = 0.186$\Rsun,
and $a = 5.65 \cdot 10^{10}$ cm (= 0.81\Rsun). From these numbers we conclude
that the companion is an M dwarf. According to the models for low-mass dwarf
stars of Dorman et al.~(\cite{dorman89}), a star of 0.14\Msun\ would have a
radius of 0.17\Rsun, fully consistent with the observations.

The above given parameters correspond to the photometric solution for
$q = 0.278$. As can be seen from Fig.~\ref{sdb_hrd}, the position of
\object{HS~0705+6700} in the HRD is well compatible with those of other EHB 
stars as well as with EHB evolutionary tracks of stars with corresponding 
masses. Considering the alternate photometric solution for $q = 0.35$,
one would have to assume a lower primary mass of say $M_1 = 0.3$\Msun. 
In principal, such mass could result as end product of the common envelope 
evolution. The HRD position of the primary might then also match a track
leading to a helium white dwarf. However, as already mentioned earlier,
the reduced mass of the M dwarf would require a smaller radius for this 
star, hardly compatible with the light curve solution.

Given the most probable parameters and system configuration of
\object{HS~0705+6700} (Fig.~\ref{meri}), the further evolution might lead 
to the formation of a cataclysmic system with a period below the period gap.
The process of angular momentum loss by gravitational radiation would require
about $10^9$ years for the binary components to merge. Should the radius of
the sdB star along its track towards the white dwarf cooling sequence,
however, increase sufficiently to reach its limiting lobe, another episode of
rapid mass transfer and loss of mass and angular momentum, perhaps even
another common envelope phase could occur within a much shorter time scale.
Yet, such prospects should be subject of detailed evolutionary calculations
lying beyond the scope of this paper.
 
\begin{table*}
\begin{center}
\caption{System parameters of three similar eclipsing sdB binaries}
\label{3sys}
\begin{tabular}{cccc}
\hline\noalign{\smallskip}
      & \object{HW~Vir}  & \object{PG~1336-018} & \object{HS~0705+6700}\\[1mm]
\hline\noalign{\smallskip}
Author   & Wood et al. 1993 & Kilkenny et al. 1998 & this paper \\
data     & $UBVR$           & $UVR$                & $BR$       \\
$P$      & $2^\mathrm h\,48^\mathrm m$ & $2^\mathrm h\,25^\mathrm m$ &
                                         $2^\mathrm h\,18^\mathrm m$  \\
$q = M_2/M_1$ & $\sim 0.3$     & 0.3                 & 0.278          \\
$i$      & $80\degr.6$         & $81\degr$           & $84\degr.4$    \\
$K_1$    & 87.9\kms            & 78\kms              & 85.8\kms       \\
$T_1$    & $\sim 33\,000$~K    & $33\,000$~K         & $29\,600$~K    \\
$T_2$    & $\sim  3\,700$~K    & $\sim  3\,000$~K    & $ 2\,900$~K    \\
$\log g_1$ & 5.64              & 5.7                 & 5.40           \\
$M_1$    & 0.54\Msun           & $\sim 0.50$\Msun    & 0.483\Msun     \\
$M_2$    & 0.18\Msun           & 0.17\Msun           & 0.134\Msun     \\
$R_1$    & 0.183\Rsun          & 0.165\Rsun          & 0.230\Rsun     \\
$R_2$    & 0.188\Rsun          & 0.175\Rsun          & 0.186\Rsun     \\
$a$      & 0.89\Rsun           & 0.79\Rsun           & 0.81\Rsun      \\[1mm]
\hline
\end{tabular}
\end{center}
\end{table*}
 
\subsection{Similar systems}
\label{simsys}

HS~0705+6700 is only the third eclipsing binary known to consist of an sdB
star and an M dwarf companion. An orbital period of about 0.096 days places
\object{HS~0705+6700} in a period range, which coincides with the period gap
of cataclysmic variables. There are two other surprisingly similar systems --
\object{HW~Vir} (P $\approx$ 0.117~d) and \object{PG~1336-018} (P $\approx$ 
0.101~d, ) -- which have equally short periods within the CV period gap
between about 2 and 3 hours (e.g., Shafter \cite{shafter92}). A non-eclipsing
system (\object{PG~1017-086}) with an even shorter period (P = 0.073~d) than
\object{HS~0705+6700} was discovered recently by Maxted et 
al.~(\cite{maxted01b}). The shrinkage of the orbit from an originally much 
wider binary with a red giant component to a component separation of less 
than one solar radius can be explained by a common envelope stage, during
which a large fraction of orbital angular momentum is lost. The system 
parameters of the three related sdB binaries are compared in Table~\ref{3sys}.

\begin{figure}
\includegraphics[width=8.8cm]{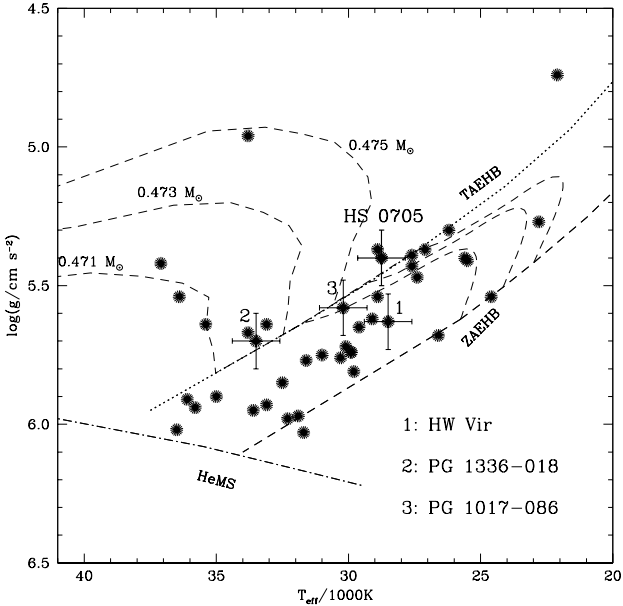}
\caption[]{Comparison of the position of \object{HS~0705+6700} 
           in the (T$_{\rm eff}$, $\log g$) diagram to that of \object{HW~Vir},
           \object{PG~1336-018} and \object{PG~1017-086}, EHB stars studied by
           Maxted et al.~(\cite{maxted01a}) and to evolutionary EHB tracks of
           Dorman et al.~(\cite{dorman93}); the tracks are labeled with 
           the stellar mass.}
\label{sdb_hrd}
\end{figure}

In Fig.~\ref{sdb_hrd} we compare the position of the four systems with 
a sample of sdB stars recently analyzed by Maxted et al.~(\cite{maxted01a})
and with calculations for EHB evolution. As can be seen all four stars lie in
the same region as the other sdB stars on the EHB band, and are therefore 
core helium burning stars with very small inert hydrogen-rich envelopes.
In order to evolve to their present configuration the systems must have
gone through a common envelope phase, when the primary was near the tip
of the red giant branch. Both the very short periods as well as the extremely
thin hydrogen layers on top of the helium cores are conceivably explained by
such common envelope evolution connected with extensive mass loss.

In the case of \object{HS~0705+6700} and the mentioned related systems the
evolutionary scenario through which the sdB stars were formed is most probably
the same. This raises again the question whether all sdB stars are members of
close binary systems and originate from the same common envelope evolution.
To improve the statistical significance of such a hypothesis, it will
therefore be of major importance to search for and analyze a larger number
of sdB candidates with special emphasis of their possible binarity.

\section{Conclusions} 
\label{concl}

The sdB star \object{HS~0705+6700} was discovered to be an eclipsing binary
with an orbital period of 8263.87~s. The analysis of the light curve revealed
that the mass ratio of the system is $q = 0.28$ and the inclination 84$\fdg4$.
The companion does not contribute to the optical light of the system, except
for the light reflected from the hemisphere facing the sdB star. The
semi-amplitude of the radial velocity curve $K_1 = 85.8$\kms\ and a mass
function of $f(m) = 0.00626$\Msun\ were derived. Accurate absolute masses and 
radii of the sdB primary and the M dwarf companion could be determined. The
spectroscopic analysis of the sdB component resulted in
$T_{\rm eff} = 28\,800 \pm 900$~K, $\log g = 5.40 \pm 0.1$, and
$\log (n_{\rm He}/n_{\rm H}) = -2.68 \pm 0.15$.

The position of the sdB star in the $\log g$ - $T_{\rm eff}$ diagram
coincides with the domain of many other EHB stars. The current location
can be matched by the evolutionary track for such a star with corresponding
mass. The photometric value of the mass ratio of $q = 0.28$ and the 
respective mass of the secondary are in good agreement with the observed 
mass function and with the predicted mass-radius relation for low-mass dwarf
stars. Based on these arguments an alternate photometric $q$ value of 0.35 
appears less probable.

The further evolution of the system might lead to the formation of a 
cataclysmic system with a period below the CV period gap.

\begin{acknowledgements}
We thank E.M. Pauli and C. Karl for their assistance with the Twin 
observations and their reduction. Thanks are also due to L. Yungelson for
helpful discussions on the evolutionary stage.
We acknowledge valuable suggestions made by the referee, P. Maxted, to 
improve the manuscript.

The time-series data from the Nordic Optical Telescope have been taken
using ALFOSC, which is owned by the Instituto de Astrofisica de Andalucia (IAA)
and operated at the Nordic Optical Telescope under agreement between IAA and
the NBIfAFG of the Astronomical Observatory of Copenhagen. 

Part of this research has been made possible through DFG research
grants We~1312/23-1 (JD) and Dr~281/13-1 (SS). The observations at Calar
Alto were supported through DFG travel grant We~1312/28-1.

\end{acknowledgements}

\end{document}